\documentclass[12pt]{iopart}
\usepackage{graphicx}

\begin{document}

\title{Flow in ultra-relativistic heavy ion collisions}

\author{Fabrice Reti\`ere 
\footnote[3]{To
whom correspondence should be addressed (fgretiere@lbl.gov)}
}

\address{Lawrence Berkeley National Laboratory,
Berkeley CA 94720, 
USA}

\begin{abstract}

Flow develops in ultra-relativistic heavy-ion collisions via re-interactions among partons or/and hadrons. Characterizing flow is a crucial step towards understanding the formation of partonic matter. We review new measurements on anisotropic flow, including $v_2$ at large rapidity, $v_1$, $v_4$, and new ways of addressing non-flow issues. We then show the self-consistency of the data and characterize flow of pions, kaons, protons and $\Xi$ within the Blast Wave parameterization. Finally, we show that the data currently available suggests that flow develops at both partonic and hadronic stages. 

\end{abstract}




\section{Introduction}

Initially, the collisions between gold nuclei release a large amount of energy over a volume on the order of a few 10$^{th}$ to a few 100$^{th}$ of fm$^3$. Converting this energy density into interacting particles leads to flow directed towards the vacuum outside the collision region.  The nature of the interacting particles remains to be determined. However, the initial state is so dense that the degrees of freedom are not ordinary hadrons as they would completely overlap each others. On the other hand, hadrons are the relevant degree of freedom of the system final state.  The transition from the initial to the final state may include the formation of thermalized matter made of quarks and gluons. As flow builds up through the whole evolution of the system, understanding its properties may help disentangle partonic collective behavior from hadronic collective behavior.

Flow means not only that the system expands collectively but also that local equilibrium is achieved. Even though, we cannot prove such equilibrium, we will use the world flow in place of collective expansion. Such flow manifests itself in several different ways, leading to the following observables:
\begin{itemize}
\item {\it Transverse momentum distribution}. 
 As particles push each other away from the hot center, they acquire a flow velocity pointing towards the vacuum, which increases their momentum; the larger the particle mass, the larger the momentum increase. The shape of the transverse momentum distribution is distorted due to flow, which may be quantified using flow motivated parameterizations.
\item {\it Anisotropic flow}. 
When the gold nuclei do not collide head-on the initial energy density is not azimuthally isotropic. As flow follows the energy density gradient, it will be stronger along the short overlap direction (so, called in-plane) than along the long ovelap direction (out-of-plane), leading to an azimuthal momentum anisotropy of the particle emission.  This anisotropy is quantified by decomposing the particle azymuthal momentum distribution in Fourier coeficients such as
$dN/dp_T/d\phi = dN/dp_T (1+2 v_1 cos(\phi) + 2 v_2 cos(2\phi) + 2 v_4 cos(4\phi) + ...)$~\cite{ArtSergeiPaper}. 
\item {\it Pion source size (HBT)}. 
The two-particle interferometry technique (HBT) probes the distance between pions having roughly the same momentum.  The source size measured by two-pion interferometry may thus be used to study the space-momentum correlations that are a corollary to flow. The relationship between flow and two-particle correlation may be investigated by flow motivated parameterizations. 
\item {\it Shift between the average emission points} of different particle species probed by non-identical two-particles correlation. It has been shown~\cite{NonId,BWPaper}  that the shift between the average emission point of different particle species may be explained by flow. 
\end{itemize} 

In these proceedings, we review the latest results on anisotropic flow. Then, we investigate the self-consistency of the flow observables in the light of the latest RHIC data. We study whether or not the currently available data can discriminate between flow built up through interactions among partons, hadrons or both.

\section{New results on anisotropic flow}

Elliptic flow quantified by $v_2$ has been extensively studied at SPS and RHIC. So far the RHIC data mainly focused on mid-rapidity (Y $<$ 0.5). New data on $v_2$ at large rapidity were presented at this conference by the PHOBOS~\cite{PhobosVQM} and STAR~\cite{ArtQM} collaborations. $v_2$ decreases with increasing rapidity, which follows the particle multiplicity. Studying quantitatively how $v_2$ scales with multiplicity by varying rapidity but also centrality and collision energy may shed light on the mechanism that drives flow.  Because the initial pressure gradients are fully determined by the centrality selection, varying the initial energy density by varying the rapidity and collision energy may provide access to the equation of state, as long as the system remains thermalized. For a detailed discussion, see~\cite{HeinzQM}.

New data on $v_1$ were shown at this conference by the PHENIX, PHOBOS~\cite{PhobosVQM}  and STAR~\cite{AihongQM} collaborations. While $v_1$ is small at mid-rapidity it becomes significant above a rapidity of 2. At the rapidity where $v_1$ is currently measured,  the particle yield is still dominated by pions~\cite{BRAHMSYieldQM}. Thus, the measured $v_1$ arises because nucleons from the colliding gold nuclei prevent the pions from being emitted in one direction. It would be interesting to verify this hypothesis measuring $v_1$ for identified particles as nucleons would then exhibit a different $v_1$.

The STAR collaboration has reported that there are momentum anisotropies beyond the first and second order coefficients~\cite{ArtQM,STARHighVPaper}. The $v_4$ coefficient rises to a few percent at mid-rapidity. It provides new constraints to models. So far, models and parameterizations fall short of describing the data without significant modifications. 

While $v_2$ is called elliptic flow, it may arise from other phenomena than flow such as jet or opaque sources. Disentangling flow contribution in $v_2$ from the other contributions relies on the fact that flow affects all the emitted particles while jets affect only a limited number of particles and their behavior is likely independent of the rest of the system. Using p-p collisions as a baseline to quantify the effect of jets, it has been shown that the anisotropy measured in Au-Au collisions at RHIC are much larger than the expectation from superposition of jets, proving that $v_2$ is generated by flow~\cite{AihongQM}. The same issue may also be addressed using the Lee-Yang zero technique~\cite{Borghini}.

\section{Flow systematics}

As discussed in the introduction, flow manifests itself in several different observables, which must be reproduced by models and parameterizations. Partonic cascade models~\cite{PartonicCascade}, hadronic cascade models~\cite{Zabrodin} and hydrodynamic models~\cite{PKolbUHeinz}, all reproduce a subset of the data. However, 
the wealth of new data coming from RHIC, at different centralities, rapidities, for many particle species, make very challenging the task of assessing whether or not a model provides a self-consistent description of flow. Releasing the model calculation code would provide the experimentalist a mean to investigate such self-consistency issues.

Data are also interpreted using parameterizations. Parameterizations are used to assess the self-consistency of the flow related observables, extract quantitative parameters and investigate specific questions, e.g. are the particle yields frozen out at the same temperature as the particle momenta. The Buda-Lund~\cite{BudaLund} and "Krakow "~\cite{Krakow} parameterizations are based on Hubble flow and a common freeze-out temperature determining both particle yields and momenta.  They reproduce well transverse mass spectra, $v_2$, and HBT data at RHIC. The Blast Wave parameterization decouples flow in the transverse plane and the beam direction~\cite{BWPaper}. A linear flow profile in transverse rapidity is used. An example of the very good description of the data provided by the Blast Wave is shown in figure 1. The Blast Wave parameterization used in these proceedings does not include resonance feed-down, which is an effect that must be present, but was found to have a minor effect on spectra~\cite{BWResonances}. The Blast Wave parameterization is also in agreement with the shift between the average emission points of pions and kaons and pions and protons measured by non-identical two-particle correlations~\cite{NonId}. Thus, the success of the parameterizations shows that the flow-related observables can be self-consistently tied together.

\begin{figure}
\begin{minipage}[t]{0.49\textwidth}
\includegraphics[width=\textwidth]{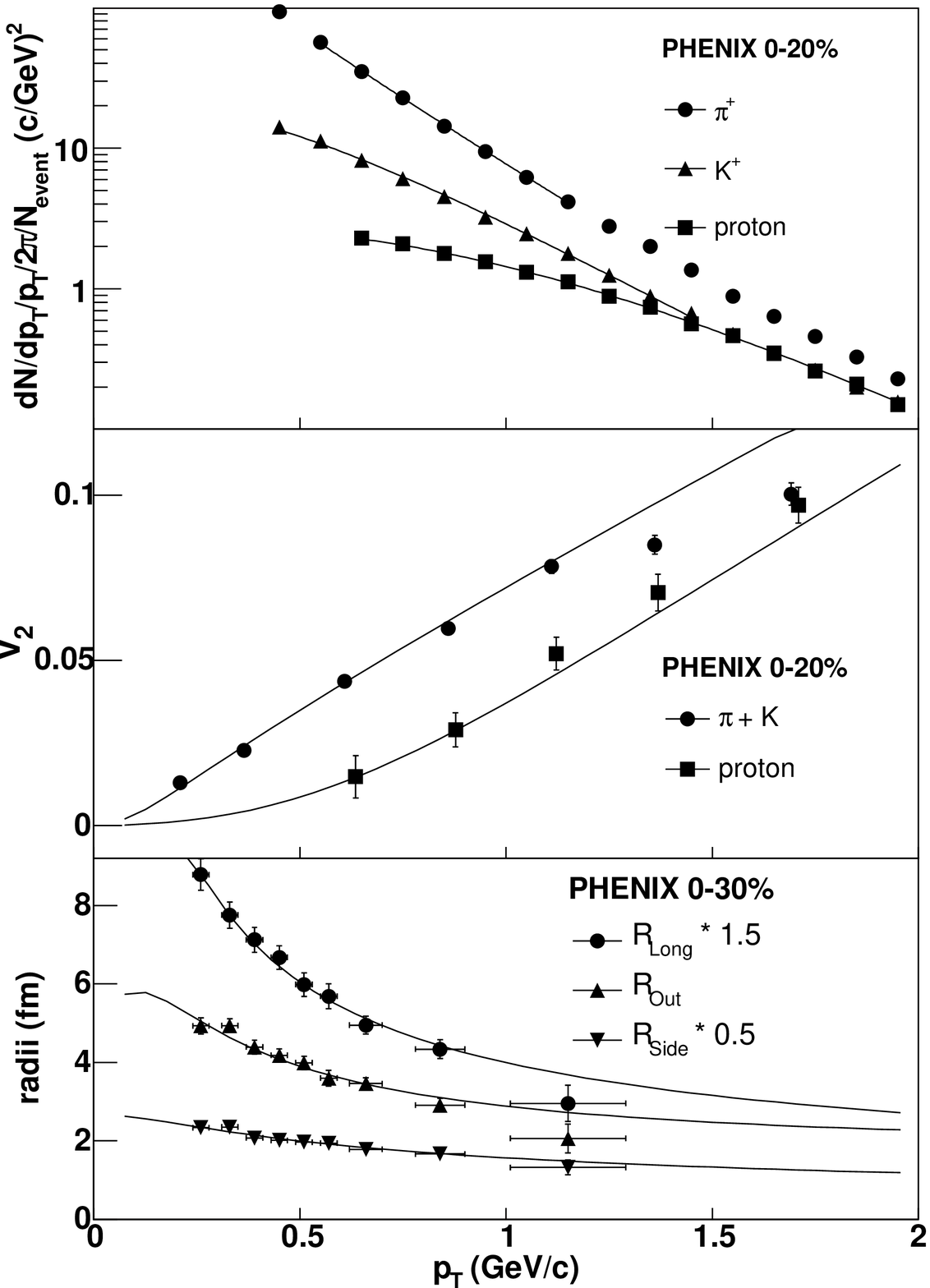}
\caption{Blast wave fit to PHENIX spectra, $v_2$, and HBT data~\cite{PHENIXPublished}.
}
\label{Fig1}
\end{minipage}
\begin{minipage}[t]{0.49\textwidth}
\includegraphics[width=\textwidth]{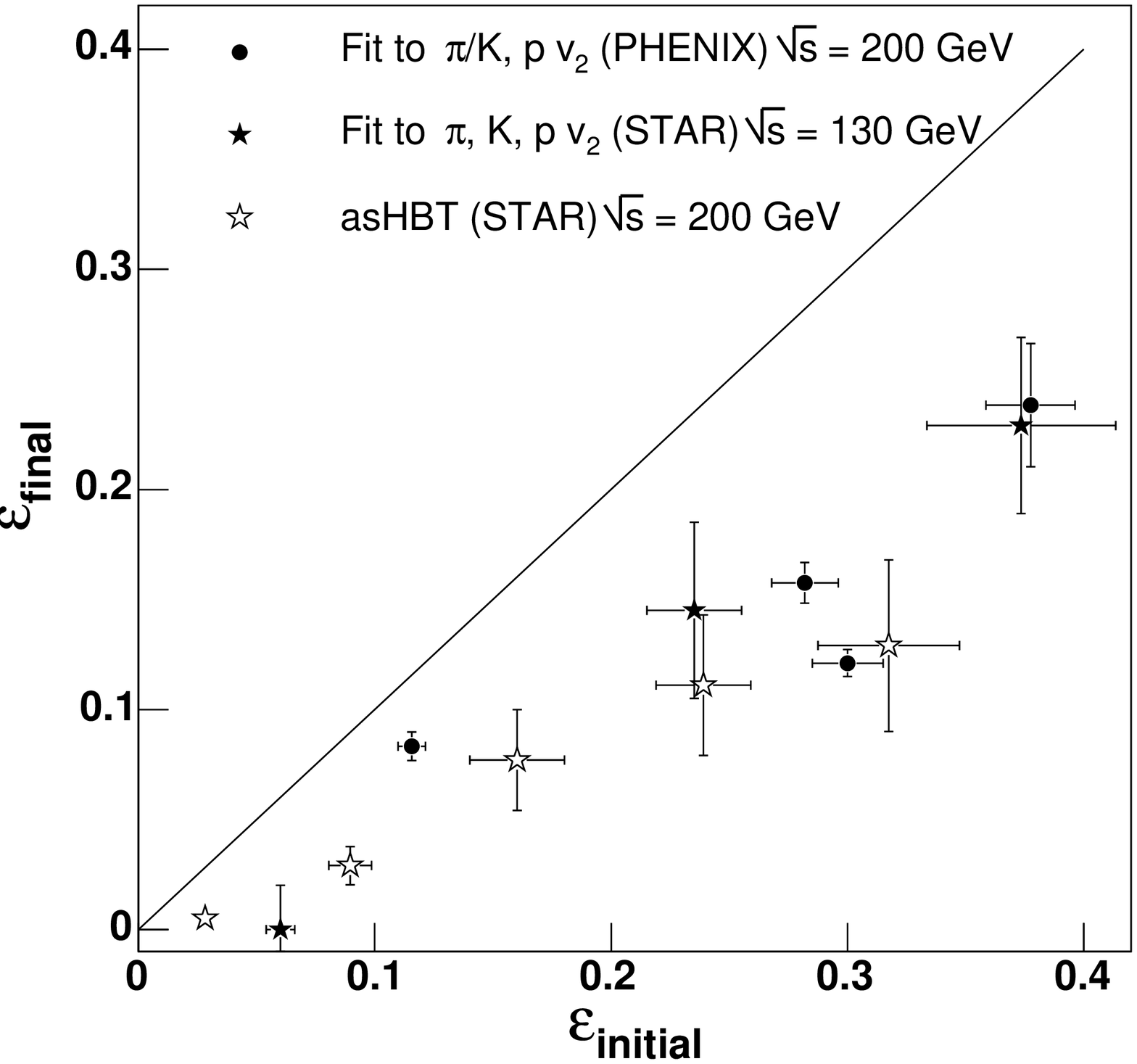}
\caption{Eccentricity extracted from fit to $v_2$ in Au-Au collision at $\sqrt{s_{NN}}=200 $ GeV~\cite{PHENIXPublished} and  $\sqrt{s_{NN}}=130 $ GeV~\cite{BWPaper} and from STAR azymuthaly sensitive HBT~\cite{STARHBT}.
}
\label{Fig2}
\end{minipage}
\end{figure}

Parameterizations also provide a mean to investigate how the system final state evolves varying the centrality or the collision energy. For example, the centrality dependence of the flow velocity and freeze-out temperature have been studied by the STAR collaboration~\cite{STARSpectra}.  Figure 2 shows the same kind of study, now investigating how the freeze-out eccentricity ($\epsilon = (R_{out-of-plane}^2 - R_{in-plane}^2) / (R_{out-of-plane}^2 + R_{in-plane}^2)$), extracted from a simultaneous fits to $v_2$ and spectra, evolves with centrality represented by the initial eccentricity calculated within a Glauber framework. The figure also includes eccentricities extracted from azimuthal HBT data~\cite{STARHBT}, which is nicely consistent with the ones extracted from fit to $v_2$. As expected, flow deforms the system, the expansion being larger in-plane than out-of-plane. For all centralities, this deformation does not overcome the initial aspect ratio, i.e. the system remains out-of-plane extended ($R_{out-of-plane} > R_{in-plane}$ or in other terms $\epsilon > 0$) , which suggests that freeze-out occurs early. 

\section{hadronic or/and partonic flow?}

The flow features may vary drastically depending if it builds up through interactions among partons, hadrons or both. Little is known about parton-parton interactions and how it would affect flow. Flow may be the best tool to study these interactions if hadronic flow is either in-significant or if it can be suppressed. The striking feature of hadronic flow that can be searched for in the data is its dependence on cross-sections.  Pion-hadron cross-sections are the driving forces behind hadronic flow, since pions far outnumbered the other particle species. 

When the energy available in the center of mass of the colliding particles is large, inelastic interactions dominate, modifying both the particle momenta and the relative yields of different particle species. On the other hand, close to freeze-out,  when the colliding particle momenta are less than 1-2 GeV/c, pion-hadron interactions are dominated by pseudo-elastic processes such as $\pi + p \rightarrow \Delta  \rightarrow \pi + p$, which only modify the momentum 
but not the relative yields of the interacting particles. The yield of the resonance that mediates the interaction ($\Delta$ in this example) may however be modified.  The pion-nucleon cross-section is the largest followed by $\pi + \pi$ ($\rho$), and presumably  $\pi + \Lambda$ ($\Lambda^*$), $\pi + K$ ($K^*$), $\pi + \Xi$ ($\Xi^*$) and $\pi + \Omega$. The resonances corresponding to the dominating pseudo-elastic process are indicated in parenthesis.  Thus, if flow is driven by hadronic interactions, it should have the following features:
\begin{itemize}
\item  Particles with larger cross-sections acquire more flow than particles with low cross-sections. Particles that are less likely to interact freeze-out earlier. 
\item Resonances yields may change even after the yield of $\pi$, K, p, $\Lambda$ ... is frozen out.
\end{itemize}

\begin{figure}[t]
\par\centering
\resizebox*{.7\textwidth}{!}{\includegraphics*{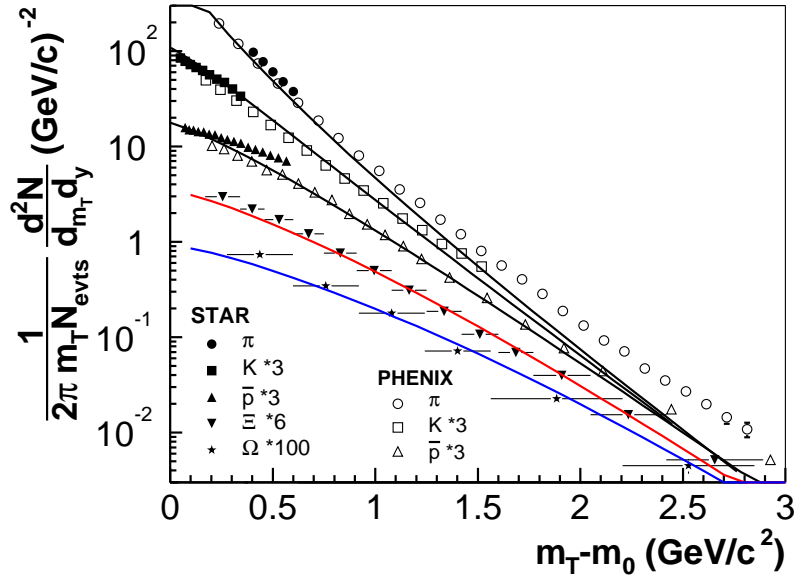}}
\caption[Fig3]{\centering \label{Fig3} 
Comparison of a hydrodynamic calculation~\cite{PKolbUHeinz} of transverse mass spectra with the data measured by the PHENIX~\cite{PHENIXPublished} and STAR collaborations~\cite{STARSpectra,JavierQM}.
 }
\end{figure}

Even though the $\pi-\Xi$ and $\pi-\Omega$ cross-sections are unknown, they are presumably significantly smaller than the $\pi-\pi$ and $\pi-p$ cross sections. Following this assumption, the significant $\Xi$ and $\Omega$ $v_2$  measured at RHIC~\cite{JavierQM} suggests that flow develops through interaction among partons in the early stage of the collision, when $v_2$ is mostly built up.

Futhermore,  if a significant flow build up at the hadronic stage, $\Xi$ and $\Omega$ transverse momentum spectra and $v_2$ should  not have the same features as $\pi$, K and protons. Figure 3 shows a comparison of measure transverse mass spectra with a hydrodynamic calculation. This calculation uses an equation of state that has two phases partonic and hadronic. However, it treats $\Xi$ and $\Omega$ exactly as any other hadrons because it assumes zero mean free path throughout the calculation independently of the type of particle involved. Figure 3 shows a qualitative agreement between the calculation and the data, which suggests that $\Xi$ and $\Omega$ freeze-out at the same temperature as $\pi$, K and protons. However, we argue that quantitatively the calculation does not describe well STAR and PHENIX data.

Fits to spectra and $v_2$ within the Blast Wave parameterization also offer a mean of investigating whether or not the same parameters can be used to describe $v_2$ and spectra of all the particle species. Figure 4 shows the best fit parameters and the one and two sigma error contours extracted from fitting spectra from NA49 at SPS (left), spectra at RHIC (middle) and $v_2$ at RHIC (right). The same contours are presented in these proceedings by the NA57 collaboration showing that $\Xi$ behave as $K_{s}^{0}$ and $\Lambda$~\cite{NA57QM}. Surprisingly, the $\Xi$  freeze-out temperature is even lower than $K_{s}^{0}$ and $\Lambda$ freeze-out temperature. On the other hand, the fit of NA49 $\Omega$ and $\phi$ spectra yield a higher temperature than the fit of $\pi$, K and $\Lambda$ spectra, but with the one $\sigma$ contours overlaping. Thus, we cannot draw definite conclusions from the data we have at hand. This issue should be resolved by fitting NA49 (including the $\Xi$) and NA57 spectra within the exact same Blast Wave framework, i.e. with the same flow profile (NA57 used a profile linear in velocity, while we used a profile linear in rapidity to fit NA49 data).

\begin{figure}[t]
\par\centering
\resizebox*{.8\textwidth}{!}{\includegraphics*{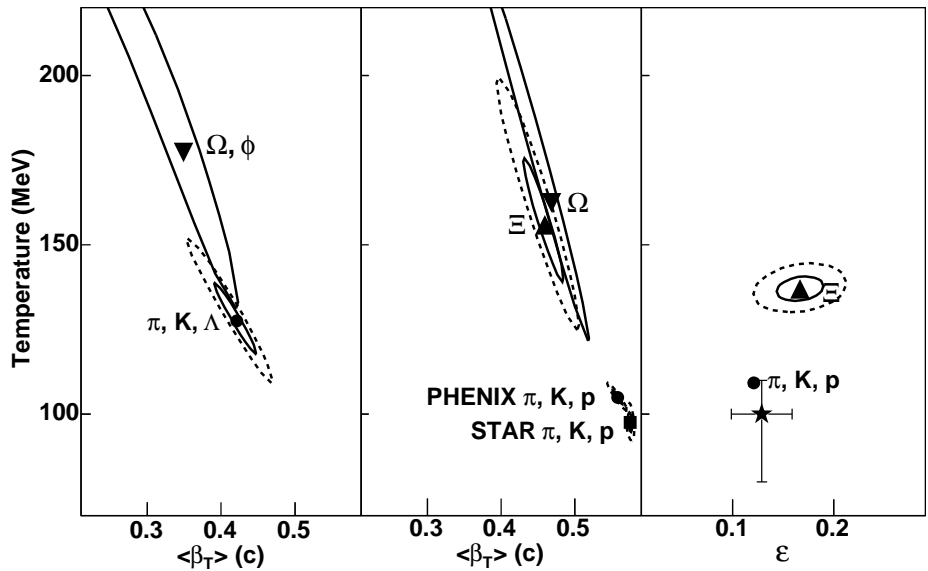}}
\caption[Fig4]{ \label{Fig4} 
Best fit parameters extracted from fits to spectra in Pb-Pb collisions at $\sqrt{s_{NN}}=17.3$ GeV (left) and fits in
Au-Au collisions at $\sqrt{s_{NN}}=200$ GeV of spectra (middle) and $v_2$ (right). Point: best fit parameter. Plain line: 1 $\sigma$ contour. Dash line: 2 $\sigma$ contour. Star symbol: calculation from STAR azymuthally sensitive HBT~\cite{STARHBT}. 
 }
\end{figure}

At RHIC energy, Blast Wave fits to STAR~\cite{STARSpectra} and PHENIX~\cite{PHENIXPublished} $\pi$, K and p spectra and $v_2$ show a freeze-out state at T $\approx$ 100 MeV, $<\beta_{T}> \approx $ 0.57, and $\epsilon \approx 0.12$.  On the other hand, fit to $\Xi$ spectra and $v_2$ yield  T $\approx$ 150 MeV, $<\beta_{T}> \approx $ 0.46, and $\epsilon \approx 0.17$, consistent with the early freeze-out picture. The statistic errors are too large to conclude for the $\Omega$. The $\Xi$ freeze-out parameters lie in between the parameters extracted from $\pi$, K and p, and the 
parameters that presumably characterize the initial state of the system, T $>$ 200 MeV, $<\beta_{T}> \approx $ 0
and $\epsilon \approx 0.3$ for minimum bias data. This suggest that $\Xi$ decouple earlier from the system than $\pi$, K and p. Assuming that the $\pi + \Xi$ and  $\pi + \Omega$ cross-sections are small, the Blast Wave fits suggests that flow has developed first within partonic matter, affecting all particles, and then within hadronic matter, affecting only particles with large hadronic cross-sections.

Resonance yields also support a scenario where significant hadronic rescattering takes place. It has been shown at this conference~\cite{Resonances} that the yields of the resonances that are involved in the pseudo-elastic interactions, do not fit the thermal expectation assuming that all particles freeze-out at the same temperature ($\approx$ 160-170 MeV).  The $\Lambda^*$ and K$^*$ yields are over-estimated while the $\Delta$ yield is under-estimated, which is consistent with a scenario where $\Lambda^*$ and $K^{*}$ decay product reinteract 
forming $\rho$ or $\Delta$, which have a larger cross-section.  In such scenario, $\rho$ and $\Delta$ yield are enhanced, while $\Lambda^*$ and K$^*$ yields are reduced. Along the same line, the $\phi$ yield fits well within the thermal freeze-out picture, because its long lifetime put its decay products out of range of any further re-interactions. 

\section{Summary and outlook}

Since the previous Quark Matter conference, a wealth of new data related flow has been gathered. An increasing number of particle species are being studied including $\Xi$, $\Omega$ and resonances such as $\Lambda^*$ or $\Delta$, shedding light on a possible interplay between partonic and hadronic flow.  In these proceedings, we have shown the new measurement of $v_1$ and the high order harmonics ($v_4$, $v_6$, ...), as well as the development of new tools coping with non-flow effects.   

Models now face the daunting task of reconciling all these data within a self-consistent framework. Such a task would be eased by releasing model codes so that experimentalists can assess how models compare to their specific data set.  While the comparison of data to models remains incomplete, parameterizations have proven useful to assess the self-consistency of the data and characterize flow. 

While the situation remains unclear at SPS energy, the Blast Wave parameterization shows that, at RHIC energy, $\Xi$ do not behave as $\pi$, K, p suggesting early freeze-out of $\Xi$ due to their low $\pi-\Xi$ cross-sections. This  suggests that $\Xi$ acquired its flow only at the partonic stage. This conclusion is based not only on fit to spectra but also on fit to $v_2$, extracting the system freeze-out eccentricity, which provides additional confidence. However, the results obtained with the Blast-Wave parameterization are challenged by a hydrodynamic calculation, which reproduces $\Xi$ and $\Omega$ spectra only if they acquire flow at both partonic and hadronic stages. On the other hand, the resonance yields are consistent with a cross-section driven hadronic stage unlike the hydrodynamic calculation, which ignores cross-sections assuming infinitely small mean free path for all particles.  No matter whether there is a significant hadronic rescattering stage, all models except~\cite{Zabrodin}, require a partonic stage to reproduce the data. Furthermore, the data at intermediate transverse momentum  ($v_2$ and spectra)  are very well reproduced within a framework assuming quark flow and hadronisation by quark coalescence~\cite{FriesQM}.  Thus, even though we do not have a proof of the partonic flow hypothesis, there are a wide variety of evidences supporting it. 

The PHENIX collaboration has shown at this conference that it was capable of addressing direct photon flow and charm flow issues~\cite{KanetaQM}, even though the statistics remained too low to draw any clear conclusions. The 2004 high statistics RHIC data hold great promise of characterizing precisely the flow of resonances,  multi-strange baryons, charm hadrons and photons, which should un-ambiguously establish whether or not flow develops in partonic matter.

I would like to thank M. Belt Tonjes, J. Bielcikova, G. Bruno, S. Esumi, M. Kaneta,  G. Roland for giving me access to their data and helping me use them. I also wish to thank   J. Castillo,  K. Filimonov, P. Jacobs, V. Koch, P. Kolb, M. Lisa, G. Odyniec, M. Oldenburg, A. Poskanzer, H.G. Ritter, K. Schweda, P. Sorensen, A. Tang, M. Van Leuwen, S. Voloshin, N. Xu, and E. Yamamoto for their valuable help in preparing the talk and these proceedings.

\end{document}